\documentclass[twocolumn]{aastex63}

\usepackage{bm}
\usepackage{url}
\usepackage{amsmath}
\usepackage{color}
\usepackage{graphicx}

\received{April 1, 2021}
\revised{May 6, 2021}
\accepted{May 28, 2021}
\submitjournal{ApJL}

\shorttitle{TWI}
\shortauthors{Ueda et al.}

\begin{document}

\title{
Thermal Wave Instability as an Origin of Gap and Ring Structures in Protoplanetary Disks 
}

\correspondingauthor{Takahiro Ueda}
\email{takahiro.ueda@nao.ac.jp}

\author[0000-0003-4902-222X]{Takahiro Ueda}
\affil{National Astronomical Observatory of Japan, Osawa 2-21-1, Mitaka, Tokyo 181-8588, Japan}

\author[0000-0002-9298-3029]{Mario Flock}
\affil{Max Planck Institute for Astronomy, K\"{o}nigstuhl 17, D-69117 Heidelberg, Germany}

\author[0000-0002-1899-8783]{Tilman Birnstiel}
\affil{University Observatory, Faculty of Physics, Ludwig-Maximilians-Universit\"{a}t M\"{u}nchen, Scheinerstr. 1, 81679 Munich, Germany}
\affil{Exzellenzcluster ORIGINS, Boltzmannstr. 2, D-85748 Garching, Germany}



\begin{abstract}
Recent millimeter and infrared observations have shown that gap and ring-like structures are common in both dust thermal emission and scattered-light of protoplanetary disks.
We investigate the impact of the so-called Thermal Wave Instability (TWI) on the millimeter and infrared scattered-light images of disks.
We perform 1+1D simulations of the TWI and confirm that the TWI operates when the disk is optically thick enough for stellar light, i.e., small-grain-to-gas mass ratio of $\gtrsim0.0001$.
The mid-plane temperature varies as the waves propagate and hence gap and ring structures can be seen in both millimeter and infrared emission.
The millimeter substructures can be observed even if the disk is fully optically thick since it is induced by the temperature variation, while density-induced substructures would disappear in the optically thick regime.
The fractional separation between TWI-induced ring and gap is $\Delta r/r \sim$ 0.2--0.4 at $\sim$ 10--50 au, which is comparable to those found by ALMA.
Due to the temperature variation, snow lines of volatile species move radially and multiple snow lines are observed even for a single species.
The wave propagation velocity is as fast as $\sim$ 0.6 ${\rm au~yr^{-1}}$, which can be potentially detected with a multi-epoch observation with a time separation of a few years.
\end{abstract}

\keywords{
protoplanetary disks --- planets and satellites: formation 
}

\section{Introduction} \label{sec:intro}
The recent ALMA observations have revealed that axisymmetric gap and ring structures are common in the dust continuum emission of protoplanetary disks \citep{Andrews18}.
The origin of these structures is still unclear, but several formation mechanisms have been proposed such as dust filtering associated with a planet-induced gap (e.g., \citealt{Pinilla+12,Dipierro+15}), snow lines \citep{Zhang+15,Okuzumi16, Pinilla17}, instability induced by dust-gas interaction \citep{TI14,TI16} or magneto-hydrodynamical effects \citep{Flock15,Ruge16,Krapp18,Riols20}.
Although most of these mechanisms produce substructures via the density variation, the observed substructures have been detected even at wavelengths where the disk is fully optically thick (e.g., \citealt{Carrasco-Gonzalez+19,Macias+21}).

The recent infrared observations have shown that the gap and ring-like structures are also common in scattered light images \citep{Avenhaus18,Garufi+18}.
Although it is unclear if the substructures in scattered light are associated with the (sub-)millimeter substructures, the prevalence of substructures in both dust and gas disk raises the questions how these substructures are created and how they are associated with the formation of planets.

\begin{figure}[ht]
\begin{center}
\includegraphics[scale=0.5]{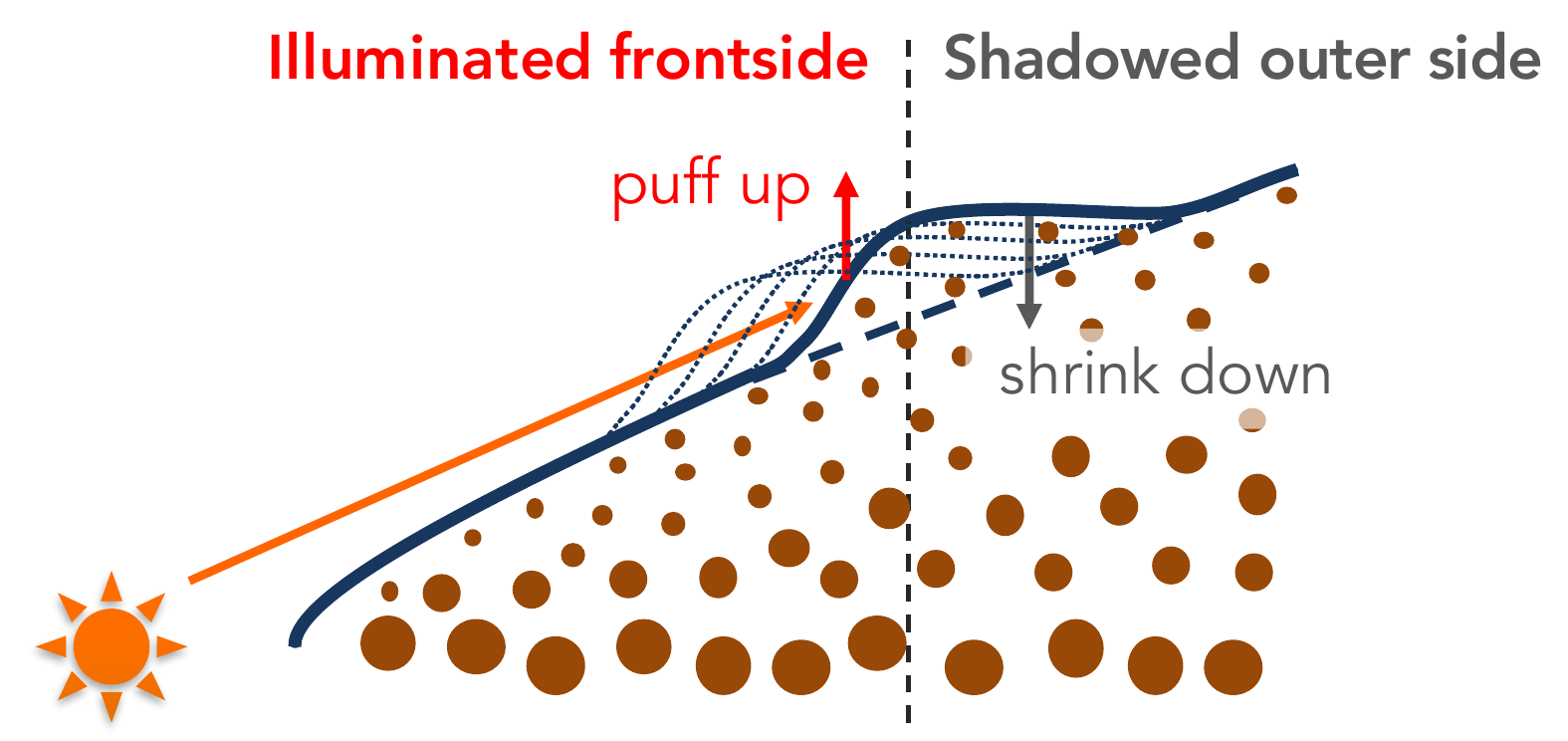}
\caption{Schematic view of the thermal wave instability.
The frontside of the perturbed disk surface receives more stellar flux and puffs up, while the shadowed outer side shrinks down (solid line).
The perturbation propagates inward as a thermal wave (dotted lines).
}
\label{fig:schematic}
\end{center}
\end{figure}

One of the potential mechanisms that creates ring and gap structures in scattered light images is the so-called Thermal Wave Instability (TWI; \citealt{DAlessio99, Dullemond00, WL08, SH12,UFO19}). 
The TWI is a physical instability induced at the surface of passively heated disks.
If the disk surface is perturbed and a small bump is generated, the illuminated frontside of the bump receives more stellar light and the shadowed outer side of the bump receives insufficient flux. 
At the illuminated side, the disk surface puffs up further as the mid-plane temperature increases, resulting in further decrease in the temperature at the backside.
Since the heating efficiency depends on the grazing angle, which is defined as the angle between the incident stellar light and the disk surface, the bump responsible of increase the grazing angle moves inward and the next bump is generated just behind the former one (Figure \ref{fig:schematic}).
In this letter, we propose the TWI as a new formation mechanism of the gap and ring structures in dust millimeter continuum emission as well as the scattered light emission.
We perform 1+1D simulations of the TWI and demonstrate that the TWI can be a potential origin of the observed gap and ring structures in both millimeter and infrared images.
The simulation model is described in Section \ref{sec:model}. 
The simulated results are shown in Section \ref{sec:results}.
Discussion and conclusion are in Section \ref{sec:discussion} and \ref{sec:summary}.

\section{Model} \label{sec:model}

\subsection{The TWI simulations}
We perform time-dependent simulations of the TWI following a model given by \citet{WL08}.
We solve a 1+1D energy equation for an axially symmetric disk.
The mid-plane temperature $T_{\rm m}$ of a passively heated disk is calculated with the energy equation
\begin{eqnarray}
\frac{\gamma+1}{2(\gamma-1)}\frac{k_{\rm B}\Sigma_{\rm g}}{\mu m_{\rm H}}\frac{\partial T_{\rm m}}{\partial t} =2(F_{\rm s}-F_{\rm m}+F_{\rm e}),
\label{eq:enegyeq}
\end{eqnarray}
where $\gamma=1.4$ is the adiabatic index, $k_{\rm B}$ is the Boltzmann constant, $\mu=2.34$ is the molecular weight of disk gas, $m_{\rm H}$ is the atomic mass unit, $\Sigma_{\rm g}$ is the gas surface density.
The three fluxes, $F_{\rm s}$, $F_{\rm m}$ and $F_{\rm e}$, represent the incident stellar flux, released energy emitted from dust and incident external flux, respectively, and are written as
\begin{eqnarray}
&&F_{\rm s}(r) = \left\{1-\exp{(-2\tau_{\rm m}(T_{\rm s})}\right\}\frac{L_{*}}{8\pi} \left\langle \frac{A_{\rm s}}{r^{2}}+\frac{4R_{*}}{3\pi r^{3}}\right\rangle, \label{eq:fs}
\\ 
&&F_{\rm m}(r) = \left\{1-\exp{(-2\tau_{\rm m}(T_{\rm m})}\right\}\sigma T_{\rm m}^{4}, \label{eq:fm}\\
&&F_{\rm e}(r) = \left\{1-\exp{(-2\tau_{\rm m}(T_{\rm m})}\right\}\sigma T_{\rm e}^{4}, \label{eq:fext}
\end{eqnarray}
where $R_{*}$ and $L_{*}$ are the stellar radius and luminosity, respectively, $T_{\rm s}$ is the temperature of the superheated dust grains at the disk surface and $\sigma$ is the Stefan-Boltzmann constant. 
Following \citet{WL08}, we set the stellar parameters as $R_{*}=2.08R_{\odot}$, $M_{*}=1M_{\odot}$, $T_{*}=4000$ K, resulting in $L_{*}=1L_{\odot}$.
The effective temperature associated with the external heating is set as $T_{\rm e}=10~{\rm K}$.
The total emitting area filling factor of superheated dust grains, $A_{\rm s}$, can be written as
\begin{eqnarray}
A_{\rm s}(r)=1-\exp\left[ - \int_{z_{\rm s}(r)}^{\infty}\kappa_{\rm s}(T_{*})\rho_{\rm d}(r,z^{\prime})dz^{\prime} \right],
\label{eq:as}
\end{eqnarray}
where $z_{\rm s}$ is the height of the disk surface where the radial optical depth for the stellar light reaches unity, $\kappa_{*}(T_{*})$ is the Planck mean opacity using the stellar temperature to determine the Black-body radiation. 
$A_{\rm s}$ is sometimes approximated with $\sin \alpha$ where $\alpha$ is the grazing angle, but we follow the more accurate determination, see Appendix C in \citet{WL08}. The dust density distribution $\rho_{\rm d}(r,z)$ is given by
\begin{eqnarray}
\rho_{\rm d}(r,z)=\frac{\Sigma_{\rm d}}{\sqrt{2\pi}h_{\rm d}}\exp{\left(- \frac{z^{2}}{2h_{\rm d}^{2}} \right)},
\label{eq:rhod}
\end{eqnarray}
where $h_{\rm d}$ is the scale height of dust disk which contributes to the opacity and is assumed to be the same as the gas scale height $h_{\rm g}=\sqrt{ k_{\rm B}T_{\rm m}r^{3}/ \mu m_{\rm H} GM_{*} }$.

Equation \eqref{eq:rhod} is based on an assumption that the disk is in hydrostatic equilibrium in the vertical direction.
This assumption is valid when the dynamical timescale $\Omega_{\rm K}^{-1}$ is shorter than the thermal timescale given by
\begin{eqnarray}
t_{\rm th}=\frac{\gamma+1}{2(\gamma-1)}\frac{k_{\rm B}\Sigma_{\rm g}}{\mu m_{\rm H}\sigma T_{\rm m}^{3}}.
\label{eq:t_th}
\end{eqnarray}
While the dynamical timescale increases as $r^{3/2}$, the thermal timescale is constant when $\Sigma_{\rm g}\propto r^{-3/2}$ and $T_{\rm m}\propto r^{-1/2}$.
Therefore, the assumption of hydrostatic would be broken at outer region ($\gtrsim$ 50 au).
However, we focus mainly on the intermediate region ($\sim$ 10--50 au) where the TWI operates and hence we assume the hydrostatic equilibrium for the entire region of the disk.

Using the dust distribution given by Equation \eqref{eq:rhod},  the vertical optical depth for the dust thermal emission is calculated as
\begin{eqnarray}
\tau_{\rm m}(T_{\rm m};r)=\int_{0}^{z_{\rm s}(r)}\kappa_{\rm m}(T_{\rm m})\rho_{\rm d}(r,z^{\prime})dz^{\prime},
\label{eq:taum}
\end{eqnarray}
where $\kappa_{\rm m}(T_{\rm m})$ is the Planck mean opacity for the dust emission.
We consider the effects of oblique radiative transfer with the similar manner used in \citet{WL08}: the angle brackets on the right-hand of Equation \eqref{eq:fs} represent a radial average of radiation emitting from surface dust with a weight of $\exp\{ -((r-r^{\prime})/z_{\rm s}(r))^{2}\}$, where $r^{\prime}$ is the radial position of the emitting surface dust.
This mimics the radial energy diffusion and stabilizes the disk against short-wavelength perturbations.
For the opacity per unit dust mass, we assume $\kappa_{*}=800~{\rm cm^{2}~g^{-1}}$ and $\kappa_{\rm m}=400~{\rm cm^{2}~g^{-1}}$.
The adopted $\kappa_{*}$ is based on the opacity of a compact dust with the MRN distribution from 0.1 to 30 ${\rm \mu m}$ and a composition of 20\% water ice, 33\% pyroxene, 40\% organics and 7\% troilite.
Even though $\kappa_{\rm m}$ depends on the dust temperature, we assume $\kappa_{\rm m}$ to be constant for the entire region of the disk for simplicity.

We use the two-dimensional cylindrical coordinate ($r,\theta$), where $r$ is the mid-plane radial distance from the central star and $\theta$ is the angle from the mid-plane.
The calculation domain ranges from 0.03 to 300 au for the radial direction and from 0 to $\pi/6$ for the $\theta$ direction, where $\theta=0$ corresponds to the mid-plane.
The radial grid is logarithmically divided into 240 bins and the $\theta$ grid is linearly divided into 360 bins.

For the disk model, we use a simple power-law gas surface density with an exponential tail, $\Sigma_{\rm g}=1700 (r/{\rm au})^{-3/2}\exp(-r/r_{\rm d})~{\rm g~cm^{-2}}$, where $r_{\rm d}$ is set to be 100 au.
The dust surface density is given as $f_{\rm d2g,s}\Sigma_{\rm g}$ where $f_{\rm d2g,s}$ represents the dust-to-gas mass ratio of small grains which contribute to the opacity.
During the simulations, the gas and dust surface density is fixed, while the temperature structure and vertical height of the disk evolve. 
This is justified when the viscous evolution timescale of the disk is longer than the thermal timescale and is valid for this setup.
For more details of the numerical procedure, we refer readers to \citet{WL08}.

\subsection{Imaging simulation}
With the temperature profile obtained from the simulations, we perform radiative transfer simulations with the Monte Carlo radiative transfer code RADMC-3D \citep{RADMC} to obtain model images.
In the simulation, we put two dust population: small grains with the maximum radius of 10 ${\rm \mu m}$ with the dust-to-gas mass ratio of $f_{\rm d2g,s}$ and large grains with the maximum radius of 1 mm with the dust-to-gas mass ratio of $0.01-f_{\rm d2g,s}$.
Each dust population has a differential size distribution with a power-law index of $-3.5$ \citep{MRN77} from the minimum dust radius of $0.1~{\rm \mu m}$.
The small grains are assumed to be well mixed with the gas. 
The scale height of the large dust grains is assumed to be 10 times lower than that of the gas to mimic the effect of vertical settling.
For the dust opacity we adopt DSHARP opacity model \citep{Birnstiel+18}.
The simulated images are convolved with a beam size of 0$\farcs$03 with assuming a distance from the Earth of 140 pc.

\section{Results} \label{sec:results}
\begin{figure*}[ht]
\begin{center}
\begin{interactive}{animation}{fig2_animation.mp4}
\includegraphics[scale=0.3]{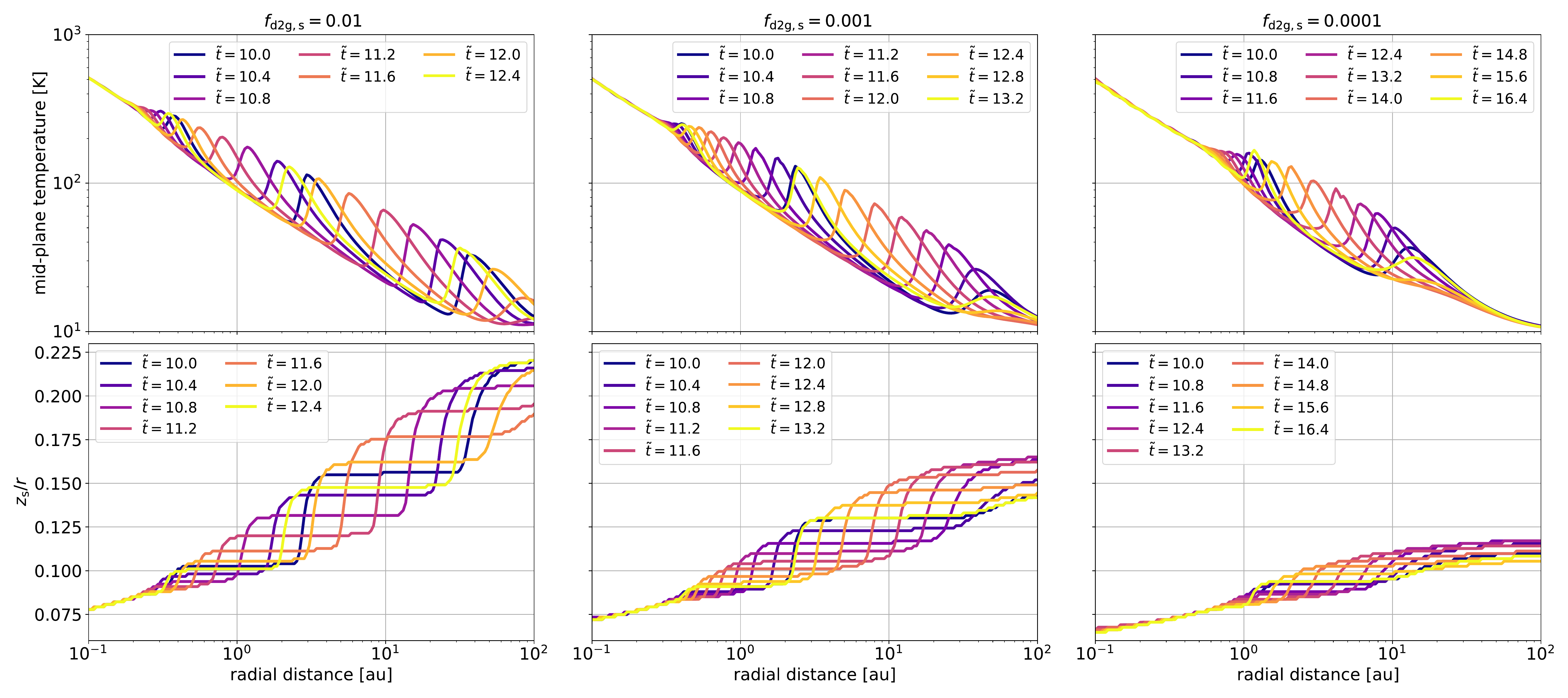}
\end{interactive}
\caption{
Time evolution of the mid-plane temperature (top) and the height of disk surface (bottom) for the model with $f_{\rm d2g,s}=0.01$ (left), 0.001 (center) and 0.0001 (right).
An animated version of this figure provides the whole time evolution of the TWI from $\tilde{t}=0$ to 19.8.
}
\label{fig:temp_zs}
\end{center}
\end{figure*}
Figure \ref{fig:temp_zs} shows the time evolution of the mid-plane temperature and the height of disk surface for the model with $f_{\rm d2g}=0.01$, 0.001 and 0.0001.
The variable $\tilde{t}$ denotes the time normalized by the initial thermal timescale $t_{\rm th,0}=41.2~{\rm yr}$.
In all models, we clearly observe the thermal wave instability.
At the illuminated frontside of the waves, the mid-plane temperature steeply increases with the radial distance.
Along with the increase in the temperature, the starlight-absorbing surface makes a steep angle to the star's rays.

the disk surface also steeply increases.
The puffed-up disk surface blocks off the stellar light and produces the shadowed outer region where $z_{\rm s}/r$ is almost constant with the radial distance.
In the shadowed region, the mid-plane temperature steeply decreases because of no direct stellar illumination.

In the model with $f_{\rm d2g,s}=0.01$, the peak of the wave located at 40 au at $\tilde{t}=10$ moves inward and reaches 10 au at $\tilde{t}=11.2$, which corresponds to a velocity of 0.625 ${\rm au~yr^{-1}}$.
The waves move more slowly at more inner region and finally vanish at $\lesssim 0.3$ au.
The propagation velocity is roughly equivalent to $r/t_{\rm th}$ when $f_{\rm d2g,s}=0.01$ but also depends on $f_{\rm d2g,s}$; the wave velocity at 10 au is 0.25 ${\rm au~yr^{-1}}$ ($\approx4\times10^{-3}v_{\rm K}$) for $f_{\rm d2g,s}=0.01$, 0.24 ${\rm au~yr^{-1}}$ ($\approx3.8\times10^{-3}v_{\rm K}$) for $f_{\rm d2g,s}=0.001$ and 0.14 ${\rm au~yr^{-1}}$ ($\approx2.2\times10^{-3}v_{\rm K}$) for $f_{\rm d2g,s}=0.0001$.
We expect that this dependence would be connected with the height of the absorption surface.
The height of the absorption surface weakly depends on the dust surface density and decreases as $z_{\rm s} \propto \sqrt{\log\Sigma_{\rm d}}$ \citep{Muto11}.
This decreasing trend in $z_{\rm s}$ can be seen in bottom panels of Figure \ref{fig:temp_zs}.
The propagation velocity is expected to be related with the rate of change in the vertical position of the absorption surface ${\partial} z_{\rm s}/{\partial} t$ which would be proportional to $z_{\rm s}$.
Since larger $f_{\rm d2g,s}$ yields larger $z_{\rm s}$, the propagation velocity would be also larger for larger $f_{\rm d2g,s}$.

\begin{figure}[ht]
\begin{center}
\includegraphics[scale=0.42]{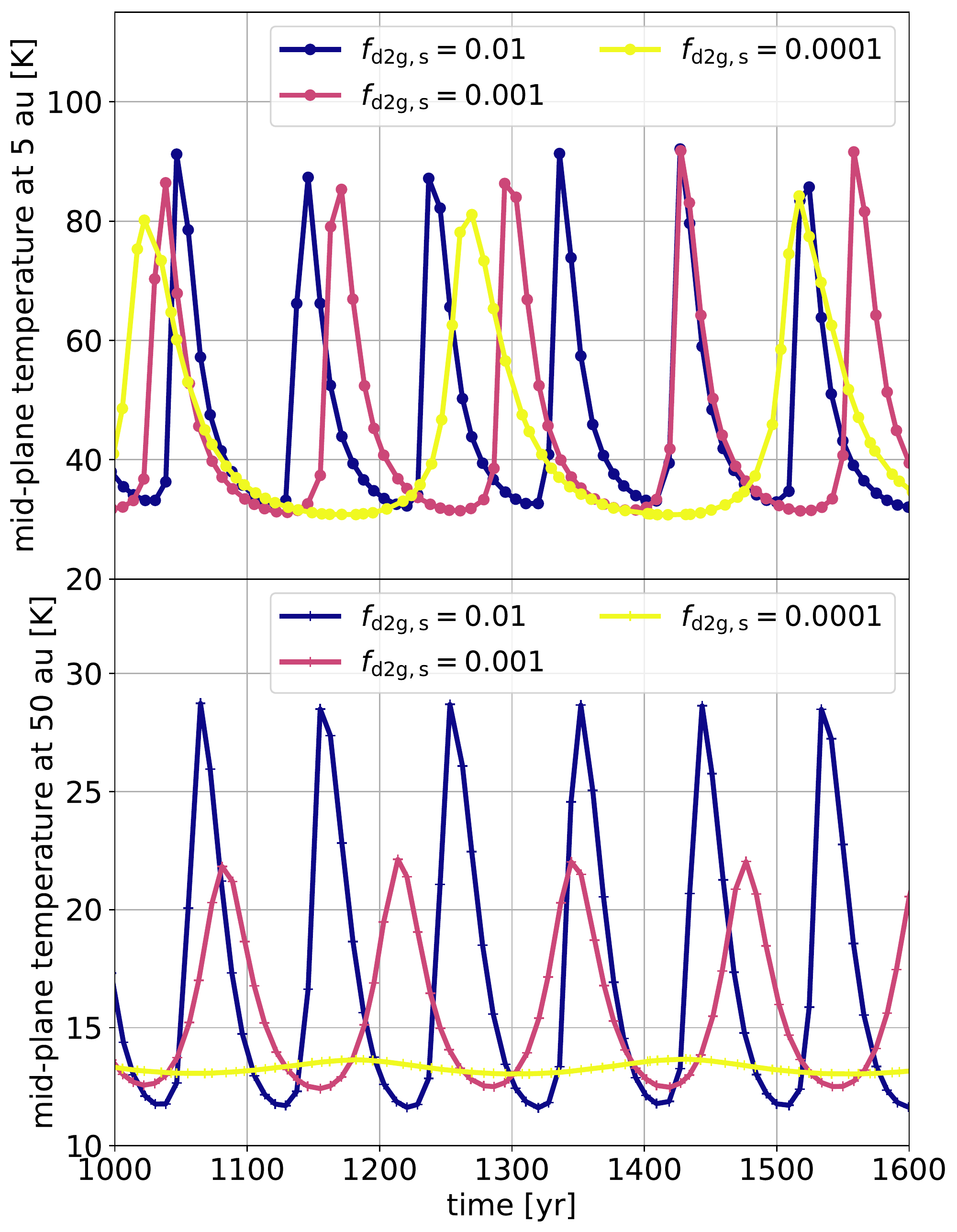}
\caption{
Time evolution of the mid-plane temperature at 5 au (top) and 50 au (bottom) for $f_{\rm d2g,s}=0.01$, 0.001 and 0.0001.
}
\label{fig:temp5au}
\end{center}
\end{figure}

For the model with $f_{\rm d2g,s}=0.01$, the mid-plane temperature oscillates with a timescale of $\sim$ 2.4$t_{\rm th,0}\approx99~{\rm yr}$.
The oscillation timescale is longer for smaller $f_{\rm d2g,s}$; $\sim$ 3.2$t_{\rm th,0}\approx132~{\rm yr}$ for $f_{\rm d2g,s}=0.001$ and $\sim$ 6.4 $t_{\rm th,0}\approx264~{\rm yr}$ for $f_{\rm d2g,s}=0.0001$.
This trend is more clearly shown in Figure \ref{fig:temp5au}.
Figure \ref{fig:temp5au} shows the mid-plane temperature at 5 and 50 au for the models with $f_{\rm d2g,s}=0.01$, 0.001 and 0.0001. 
At 5 au, the mid-plane temperature varies from $\sim$ 30 K to $\sim$ 80--90 K for all models.
The time separation of each peak is larger for smaller $f_{\rm d2g,s}$.
This is because smaller $f_{\rm d2g,s}$ makes the disk more stable for the TWI.
As the radial length-scale over which the stellar radiation is absorbed, i.e., the thickness of the absorption layer for the stellar light, is longer for smaller $f_{\rm d2g,s}$ (Appendix \ref{sec:zs}), smaller $f_{\rm d2g,s}$ is stable for longer wavelength perturbations.
Since the growth rate of the perturbation is inversely proportional to the wavelength \citep{Dullemond00}, the TWI grows more slowly for smaller $f_{\rm d2g,s}$.

The radial extent on which the TWI operates also depends on $f_{\rm d2g,s}$ and is broader for larger $f_{\rm d2g,s}$.
While for $f_{\rm d2g,s}=0.01$, the radial extent of TWI activity is between 0.3 and 100 au, this extent shrinks for $f_{\rm d2g,s}=0.0001$ to the region between 1 and 10au.
The inner edge of the unstable region is set by the irradiation from the central star with a finite radius \citep{DAlessio99}.
Since the lower $f_{\rm d2g,s}$ disk has a starlight absorbing surface closer to the mid-plane, the finite size of the central star has a bigger effect on the lower $f_{\rm d2g,s}$ disk.
It would be worth to be noted that the TWI can be also suppressed at the inner region by the accretion heating, which is ignored in our simulations \citet{WL08}.
At the outer region, the disk is stable for the TWI because the external irradiation suppresses the growth of the TWI.
Since the smaller $f_{\rm d2g,s}$ makes the disk thinner and hence cooler, the TWI can grow only at more inner region for smaller $f_{\rm d2g,s}$.
At 50 au, we clearly see that the amplitude of the temperature variation decreases with $f_{\rm d2g,s}$ and almost vanishes when $f_{\rm d2g,s}=0.0001$. 
This trend indicates that the TWI can produce the rings and gaps when the disk contains an enough amount of small dust grains.

\begin{figure*}[ht]
\begin{center}
\includegraphics[scale=0.8]{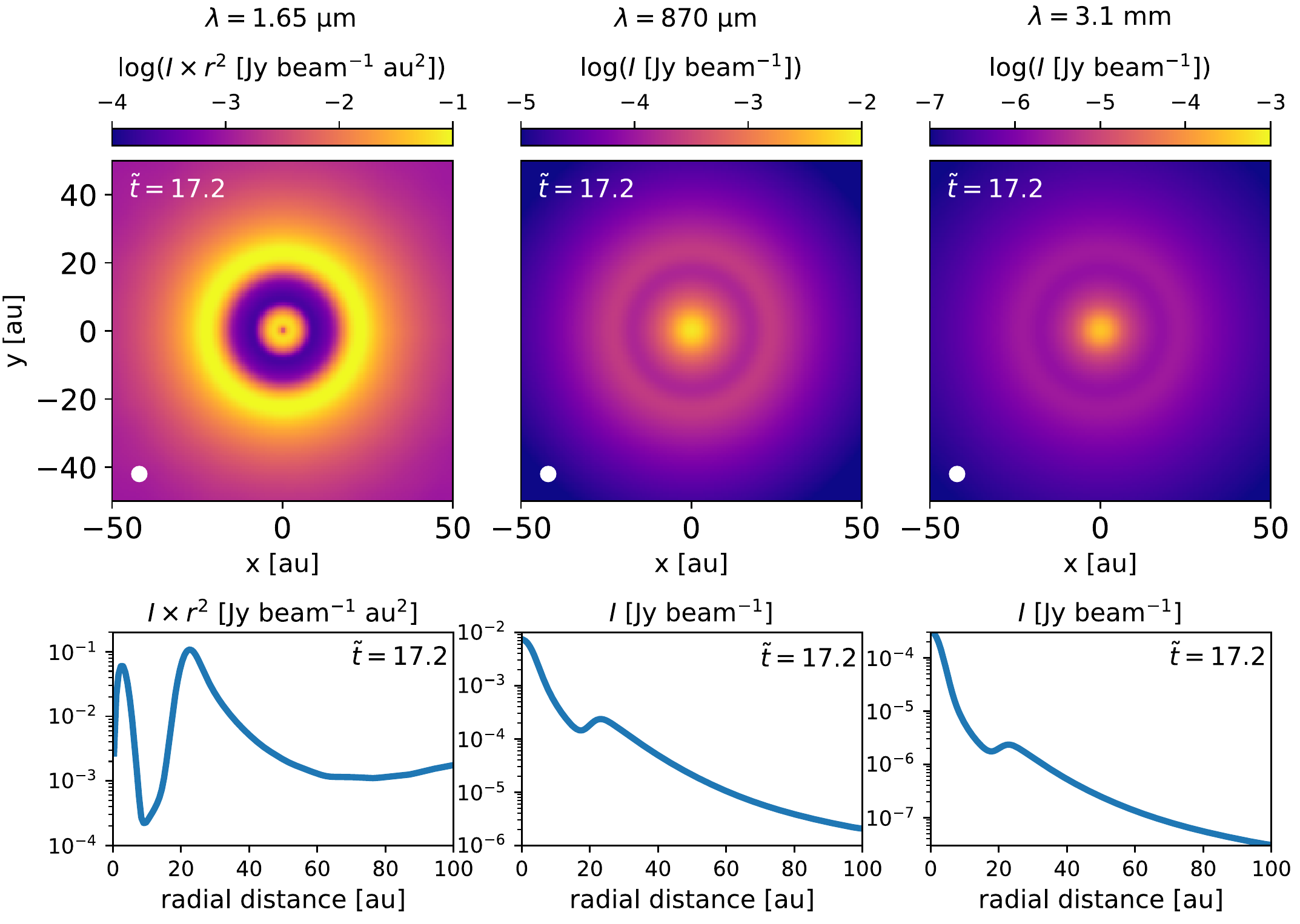}
\caption{Synthesized images of the thermal wave instability observed at $\lambda=1.65~{\rm \mu m}$ (left), $870~{\rm \mu m}$ and $3.1~{\rm mm}$ for the model with $f_{\rm d2g,s}=0.01$. 
The images are convolved with a beam size of 0$\farcs$03 which is denoted with a filled white circle at bottom left in each panel.
The simulated intensity at $\lambda=1.65~{\rm \mu m}$ is multiplied by a square of the radial distance to compensate for the stellar flux attenuation.
}
\label{fig:2dview}
\end{center}
\end{figure*}

Figure \ref{fig:2dview} shows the face-on view of the TWI-operated disk at infrared wavelength ($\lambda=1.65~{\rm \mu m}$) and millimeter wavelength ($\lambda=870~{\rm \mu m}$ and 3.1 mm). 
The simulated images are convolved with a beam size of 0$\farcs$03 which is comparable to the resolution of recent ALMA survey studies.
Both in scattered light and dust continuum images, an inner disk component with a bright ring  separated with a gap at $\sim$ 10--15 au is clearly detected.
Importantly, since the millimeter substructures are caused by the temperature variation, the substructures are visible regardless of whether the disk is optically thin or thick for its own thermal emission.
We observe the ring structure at almost the same place at all wavelengths.
The width of the gap is also almost the same at all wavelengths, but the local minimum has slight radial offset between the scattered-light and millimeter images.
This is because, the millimeter emission reflects the dust temperature which has a local minimum at the outer edge of the shadowed region, while the scattered-light images reflects the shape of the disk surface which is almost flat in the shadowed region (Figure \ref{fig:temp_zs}).
At the millimeter wavelengths, the intensity contrast between the ring and gap is $\sim$ 3, while it is more pronounced, $\sim10^{3}$, at the infrared wavelength. 
The separation of the gap and ring, $\Delta r$, is typically $\sim$ 0.2--0.4 $r_{\rm gap}$ where $r_{\rm gap}$ is the radial position of the gap, which is comparable to or slightly larger than the typical observed ones (Appendix \ref{sec:obs}).

\section{Discussion}\label{sec:discussion}

\subsection{Locations of snow lines}
Snow lines of abundant volatile species are one of the possible origins of the observed substructures in dust continuum emission \citep{Zhang+15, Okuzumi16,Pinilla17}.
Recent disk surveys have shown that the locations of the observed gap/ring structures seem to be not related with the radial locations of the snow lines  \citep{Huang+18, Long+18,vanderMarel+19}.
To estimate the positions of the snow lines in observed disks, the temperature profile is often assumed to be a simple power-law \citep{Huang+18, Long+18}, which is broken if the disk has shadows on the disk surface \citep{DDN01,BC14,UFO19}.
\begin{figure}[ht]
\begin{center}
\includegraphics[scale=0.42]{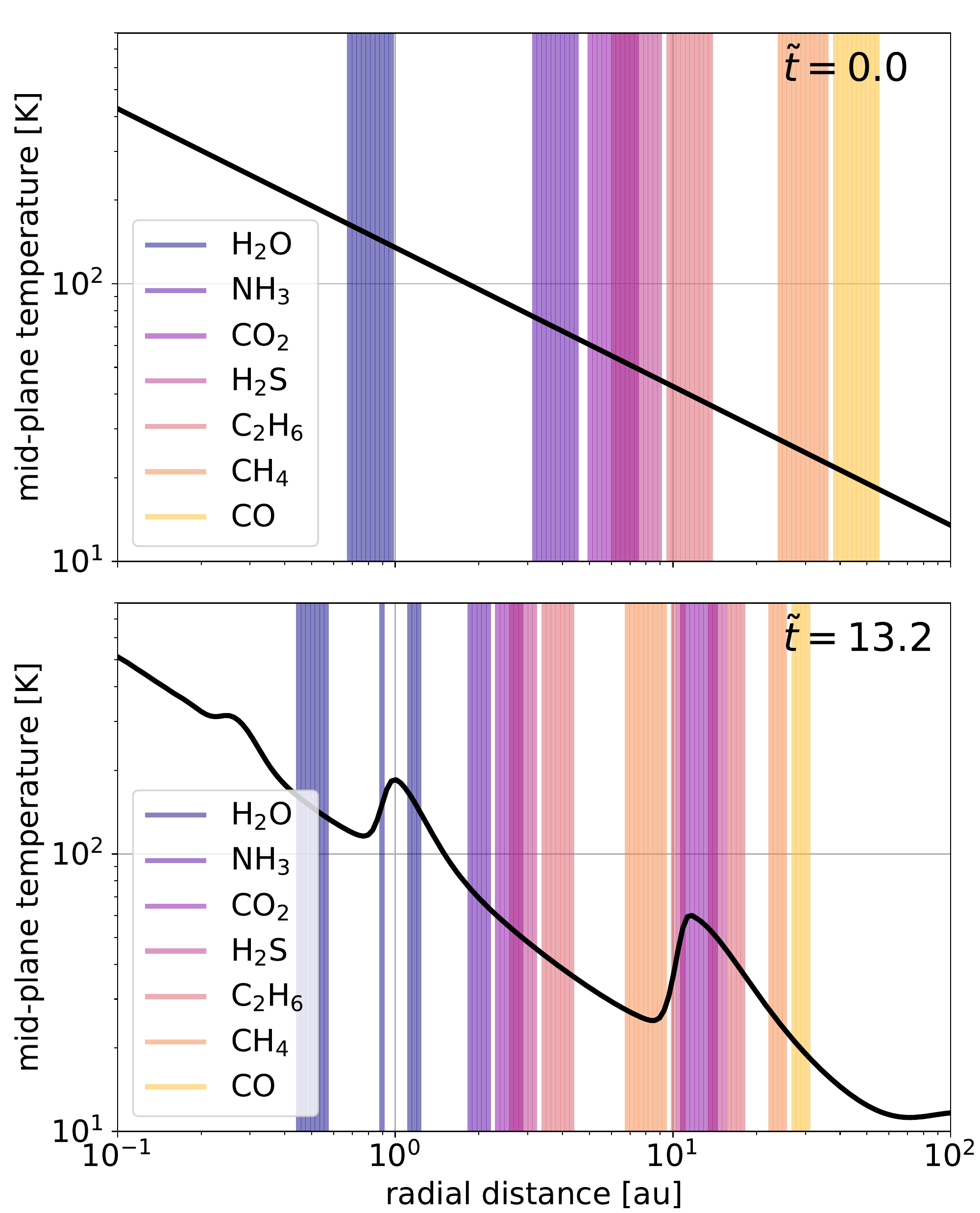}
\caption{Radial locations of the snow lines of volatile species at $\tilde{t}=0$ and 11 for $f_{\rm d2g,s}=0.01$.
}
\label{fig:SLs}
\end{center}
\end{figure}
Figure \ref{fig:SLs} shows the radial positions of snow lines of ${\rm H_{2}O}$, ${\rm NH_{3}}$, ${\rm CO_{2}}$, ${\rm H_{2}S}$, ${\rm C_{2}H_{6}}$, ${\rm CH_{4}}$ and ${\rm CO}$ at $\tilde{t}=0$ and 11.
For simplicity, we define the snow lines as the radial location where the mid-plane temperature reaches 150, 70, 55, 50, 40, 25, 20 K for ${\rm H_{2}O}$, ${\rm NH_{3}}$, ${\rm CO_{2}}$, ${\rm H_{2}S}$, ${\rm C_{2}H_{6}}$, ${\rm CH_{4}}$ and ${\rm CO}$, respectively.
We clearly see that the radial positions of snow lines move with time and multiple snow line emerges even for a single species.
This means that it is necessary to determine the disk temperature precisely when we evaluate the snow line locations.
These snow lines would induce additional ring and gap structures and some of them would overlap with the TWI-induced substructures.
It should be noted that the oscillation timescale of the TWI is much shorter than the dust radial drift timescale. 
Therefore, sintering-induced substructures would not coincide with the locations of the snow lines if the TWI is present.
However, the dust-size variation can be induced by sublimation and re-condensation, which potentially produces millimeter substructures.
Since the radial locations of the snow lines are important not only for the substructure formation but also for the chemical composition of forming planets (e.g., \citealt{Sato+16,OW19}), we should investigate how the TWI evolves in planet forming disks.

\subsection{Characteristics of TWI-induced rings and gaps}
In this section we summarize the characteristics of the TWI-induced rings/gaps and discuss how we can distinguish the TWI from the other substructure formation mechanisms.
The key points of the TWI-induced gap and ring structure are as follows.

\begin{enumerate}
\item
The intensity variation is induced by the temperature variation (i.e., variation in the gas scale height), not surface density variation.
Therefore, the substructures are visible even if the disk is fully optically thick at millimeter wavelengths.
The density-induced substructures at millimeter wavelengths would be invisible if the disk is fully optically thick.
Furthermore, even inside and within the TWI-induced gap can be filled with dust grains, while planet-induced gap would trap large grains at the outer edge of the gap and the inner region should be depleted in large dust.
Multi-wavelength millimeter observations would be helpful to distinguish the temperature-induced intensity variation from the density-induced one.

\item
The gap and ring move inward as fast as a velocity of $\sim$ 0.6 ${\rm au~yr^{-1}}$, depending on the disk optical depth and the radial position.
Multi-epoch observations with a time separation of a few years can potentially detect the radial motion of the wave.
The substructures induced by the other mechanisms would move more slowly.
For example, substructures associated with planet-induced gap would migrate with a velocity slower than the type I migration (e.g., \citealt{Kanagawa+18}).

\item
The gap and ring structure should appear both in millimeter dust continuum and infrared scattered light. 
The intensity variation is more pronounced for the scattered light.
The other mechanisms also potentially produce substructures in both since the distribution of small grains would be closely related with that of large grains if collisional fragmentation takes place (e.g., \citealt{Pinilla+15}).

\item
The fractional separation between the ring and gap is typically $\Delta r/r \sim$ 0.2--0.4. 
The other mechanisms, e.g., snow lines, might produce additional substructures between the TWI-induced ring and gap, which makes the apparent separation smaller.

\end{enumerate}

\subsection{Caveats}\label{sec:caveats}
Here we summarize our assumptions and discuss the potential effect on the TWI.
First, we assumed that the disk is in hydrostatic equilibrium in vertical direction.
This assumption would be broken in outer region ($>50$ au).
The hydrodynamical motion of the disk gas might potentially suppress the TWI at outer region, although its effect is very uncertain.
Second, we fixed the gas and dust surface densities during the TWI simulations.
Since the TWI is closely related to the disk optical depth for the stellar light, surface density evolution affect the behavior of the TWI. 
However, as shown, the evolution timescale of the TWI is very fast and much shorter than the typical viscous evolution timescale.
Therefore, the assumption of fixed gas surface density would be justified.
Even though the TWI might affect the dust surface density through the steep positive temperature gradient which traps the radially drifting dust particles, the wave propagation timescale is too fast compared to the radial drift timescale.
We compare these timescales in Appendix \ref{sec:timescales}.
Finally, we treat the disk with a 1+1D approach that mimics oblique radiative transfer but ignore the radial heat diffusion and the hydrodynamical motion of the gas.
The steep radial gradient in the temperature might induce the radial motion of the disk gas which potentially suppress the TWI.
The two-dimensional radiation hydrodynamical simulation would be necessary to investigate the TWI in more detail with taking these effect into account.

\section{Summary}\label{sec:summary}
We investigated the impact of the Thermal Wave Instability (TWI) on the millimeter and infrared emission of disks using the 1+1D simulations of disk temperature evolution.
We confirm that the TWI operates when the disk is optically thick enough for stellar light, i.e., the dust-to-gas mass ratio of small grains is $\gtrsim$ 0.0001.
The mid-plane temperature varies as the waves propagate and hence gap and ring structures can be seen in both millimeter and infrared emission even if the dust and gas surface density have no substructures.
Since the substructures are induced by the temperature variation, the millimeter substructures can be seen even if the disk is completely optically thick, while the density-induced substructures would disappear for optically thick disks.
The fractional separation between the TWI-induced ring and gap is typically $\sim$ 0.2--0.4 at $\sim$ 10--50 au, although additional substructures might form within them, e.g. by snow lines.
Due to the temperature variation, snow lines of volatile species move radially and multiple snow lines are observed even for a single species, which would affect the disk chemistry.
The wave propagation velocity is as fast as $\sim$ 0.6 ${\rm au~yr^{-1}}$, which can be potentially detected with a multi-epoch observation with a time separation of a few years.
The TWI might be stabilized by the radial energy diffusion and/or hydrodynamical motion of the disk gas, which are not fully taken into account in this work. 
The two-dimensional radiation hydrodynamical simulation should be carried out to understand how the TWI evolves in planet forming disks.

\acknowledgments
We thank an anonymous referee for the constructive feedback.
This work is supported by JSPS KAKENHI Grant Numbers JP19J01929.
T.B. acknowledges funding from the European Research Council (ERC) under the European Union's
Horizon 2020 research and innovation programme under grant agreement No 714769 and funding by the
Deutsche Forschungsgemeinschaft (DFG, German Research Foundation) under Germany's Excellence
Strategy - EXC-2094 - 390783311.
\software{RADMC-3D \citep{RADMC}}

\appendix
\section{Comparison with the observed substructures} \label{sec:obs}
In this Appendix, we compare the separation of simulated ring and gap pairs with that found by ALMA surveys.
\begin{figure}[ht]
\begin{center}
\includegraphics[scale=0.5]{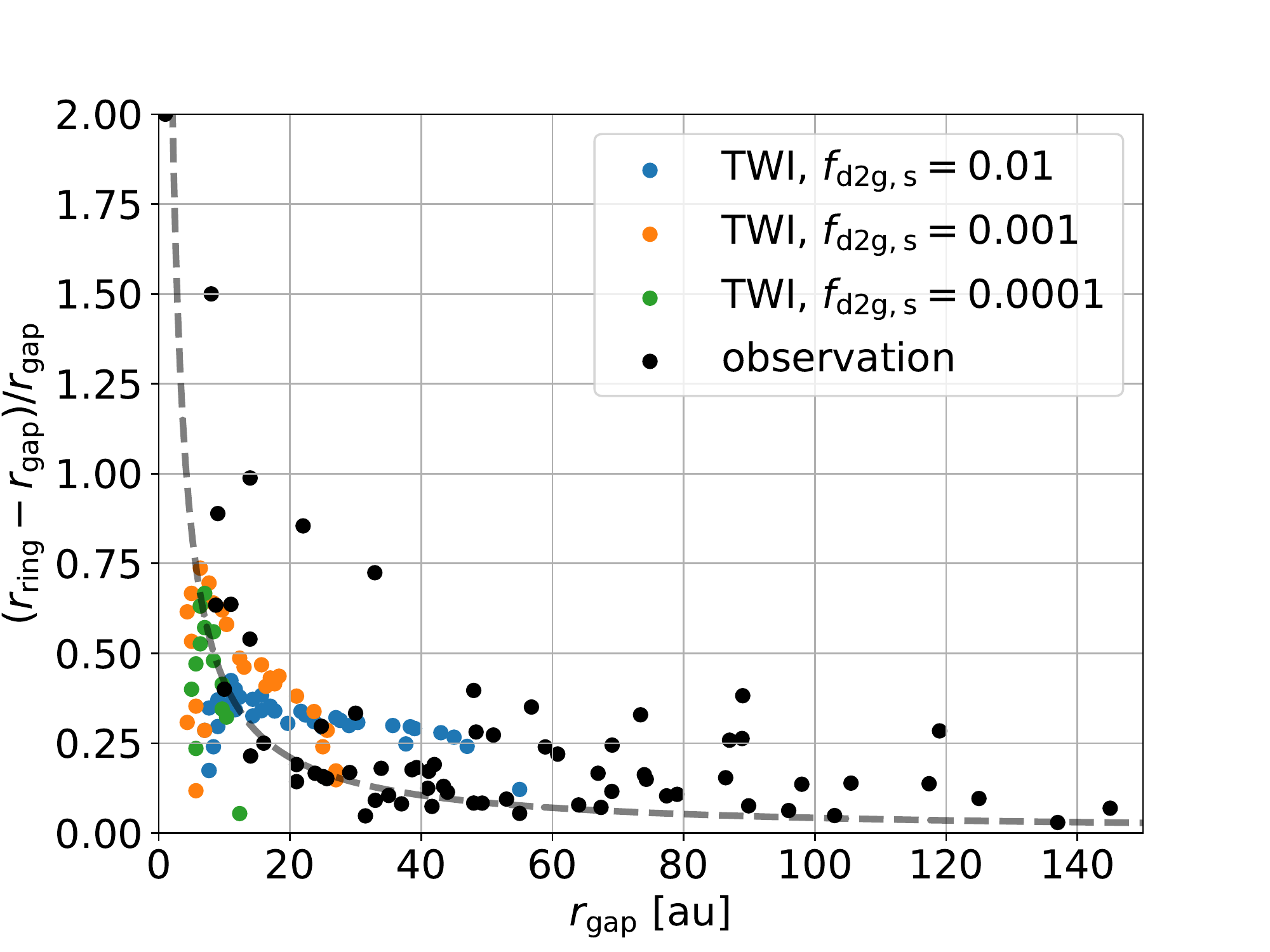}
\caption{Fractional separation of observed and simulated gap and ring pairs. 
The observational data is compiled from \citet{Huang+18} and \citet{Long+18} (see also \citealt{Andrews20}). 
The gray dashed line denotes the condition for the ring-gap separation to be equivalent to the spatial resolution corresponding to an angular resolution of 0$\farcs$03 assuming the distance from the Earth of 140 pc, i.e., $\Delta r = 4.2~{\rm au}$.
}
\label{fig:ratio}
\end{center}
\end{figure}
Figure \ref{fig:ratio} shows observed and simulated separation of ring and gap pairs $\Delta r$ normalized by the radial location of the gap $r_{\rm gap}$.
The simulated fractional separation is calculated from the model images at ALMA Band 7 ($\lambda=870~{\rm \mu m}$), while the observations are at ALMA Band 6 ($\lambda=1.3~{\rm mm}$).
Although the observing wavelength is different between the model and observation, the fractional separation of the TWI-induced ring and gap is not so sensitive to the observing wavelength since it is induced by the temperature.
The observed fractional separation is typically $<0.5$ at $>40$ au and shows an increasing trend with decreasing the radial distance.
At the intermediate disk region ($\sim$ 10-50 au), the fractional separation of the TWI-induced ring and gap pairs is $\sim$ 0.2--0.4 which is comparable or slightly higher than the observed values.
At the outer region ($\gtrsim 50 $ au), the substructures found in the observations but are not seen in the simulations with our setup.
The substructures at the outer region are potentially induced by the TWI if the disk is more massive and the central star is more luminous than our model.
At the inner region ($\lesssim10$ au), the TWI can produce ring and gap pairs with $\Delta r/r\lesssim0.75$, though most of them are not easy to be detected with a current ALMA resolution.
It should be noted that the origin of the rings and gaps would be not identical but multiple formation mechanisms would take place.
If multiple mechanisms take place in a disk, the ring and gap pairs induced by each mechanism might be overlapped.
If this is the case, the apparent separation of gap and ring pairs would be smaller than the original.

\section{The height of disk surface} \label{sec:zs}
\begin{figure}[ht]
\begin{center}
\includegraphics[scale=0.5]{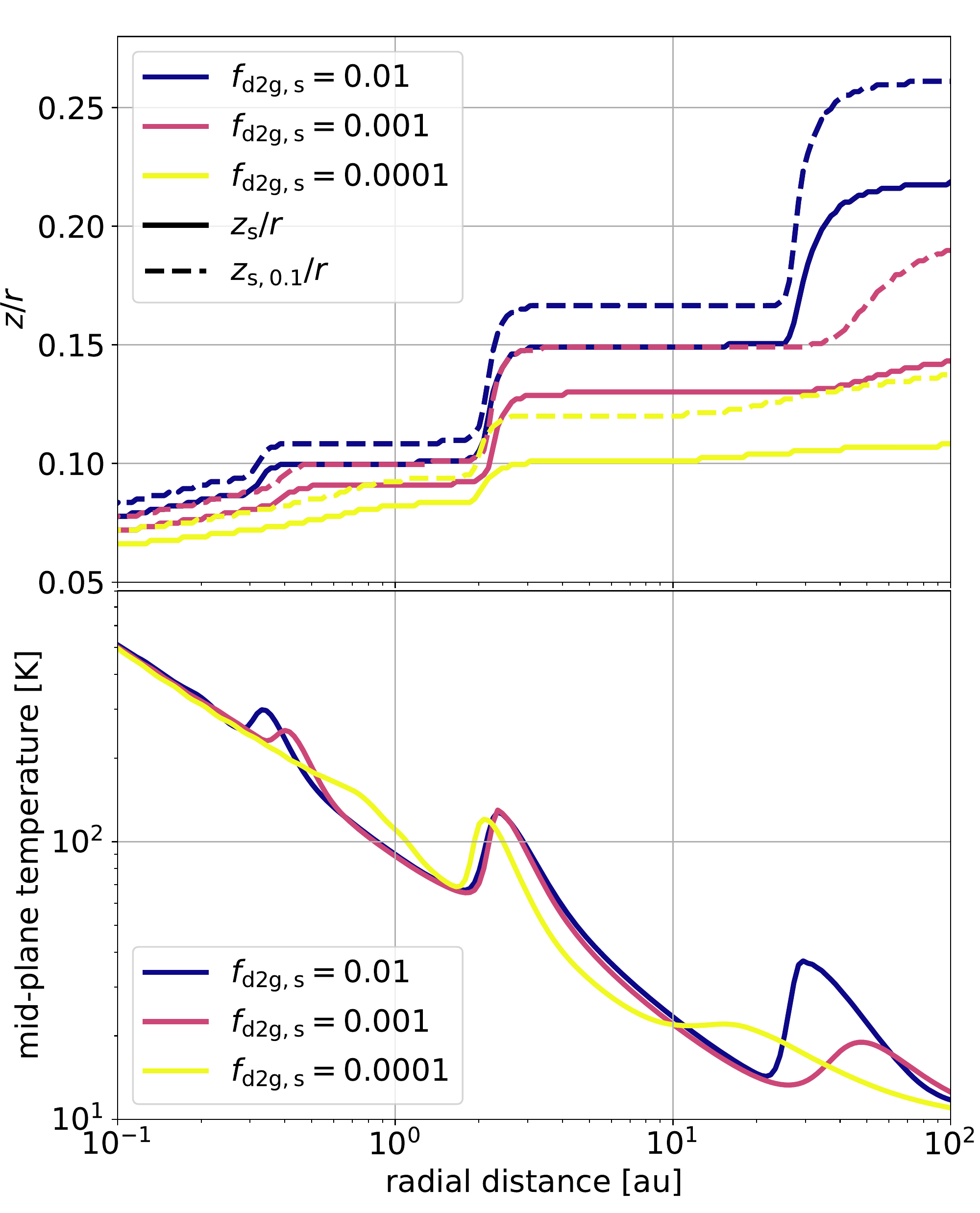}
\caption{{\it Top:} Vertical positions where the radial optical depth for the stellar light reaches 0.1 ($z_{\rm s,0.1}$; dashed lines) and 1 ($z_{\rm s}$; solid lines) for different values of $f_{\rm d2g,s}$.
{\it Bottom:} Mid-plane temperature profile for different values of $f_{\rm d2g,s}$.
}
\label{fig:zs}
\end{center}
\end{figure}

Figure \ref{fig:zs} shows the vertical locations where the radial optical depth for the stellar light reaches 0.1 ($z_{\rm s,0.1}$) and 1 ($z_{\rm s}$) for different values of $f_{\rm d2g,s}$.
For each model, we plot a time-snapshot where the wave is located at the similar location ($\sim$ 25 au) to compare the surface structure clearly. 
We see that the absorption surface is located at lower for smaller $f_{\rm d2g,s}$.
The vertical thickness of the absorption layer, i.e., the region where the optical depth ranges from $\sim$ 0.1 to 1, $\Delta z_{\rm s}$, slightly increases as $f_{\rm d2g,s}$ decreases.
As $z_{\rm s}$ decreases with $f_{\rm d2g,s}$, the thickness-to-height ratio, $\Delta z_{\rm s}/z_{\rm s}$, increases with decreasing $f_{\rm d2g,s}$ more sensitively.
Since the longest wavelength of perturbations is an order of $\sim z_{\rm s}$ and perturbations with a wavelength shorter than $\Delta z_{\rm s}$ is expected to be suppressed, larger $\Delta z_{\rm s}/z_{\rm s}$ would make the disk stable for perturbations with longer wavelengths.
This means that smaller $f_{\rm d2g,s}$ is more stable for the TWI. 
The larger $f_{\rm d2g,s}$ has a larger vertical gap in $z_{\rm s}$ at $\sim 25 $ au but the temperature variation is similar for all models.
This is because the mid-plane temperature is more sensitive to the radial gradient of the disk surface than the absolute height.

\section{Comparison of physical timescales}
\begin{figure}[ht]
\begin{center}
\includegraphics[scale=0.5]{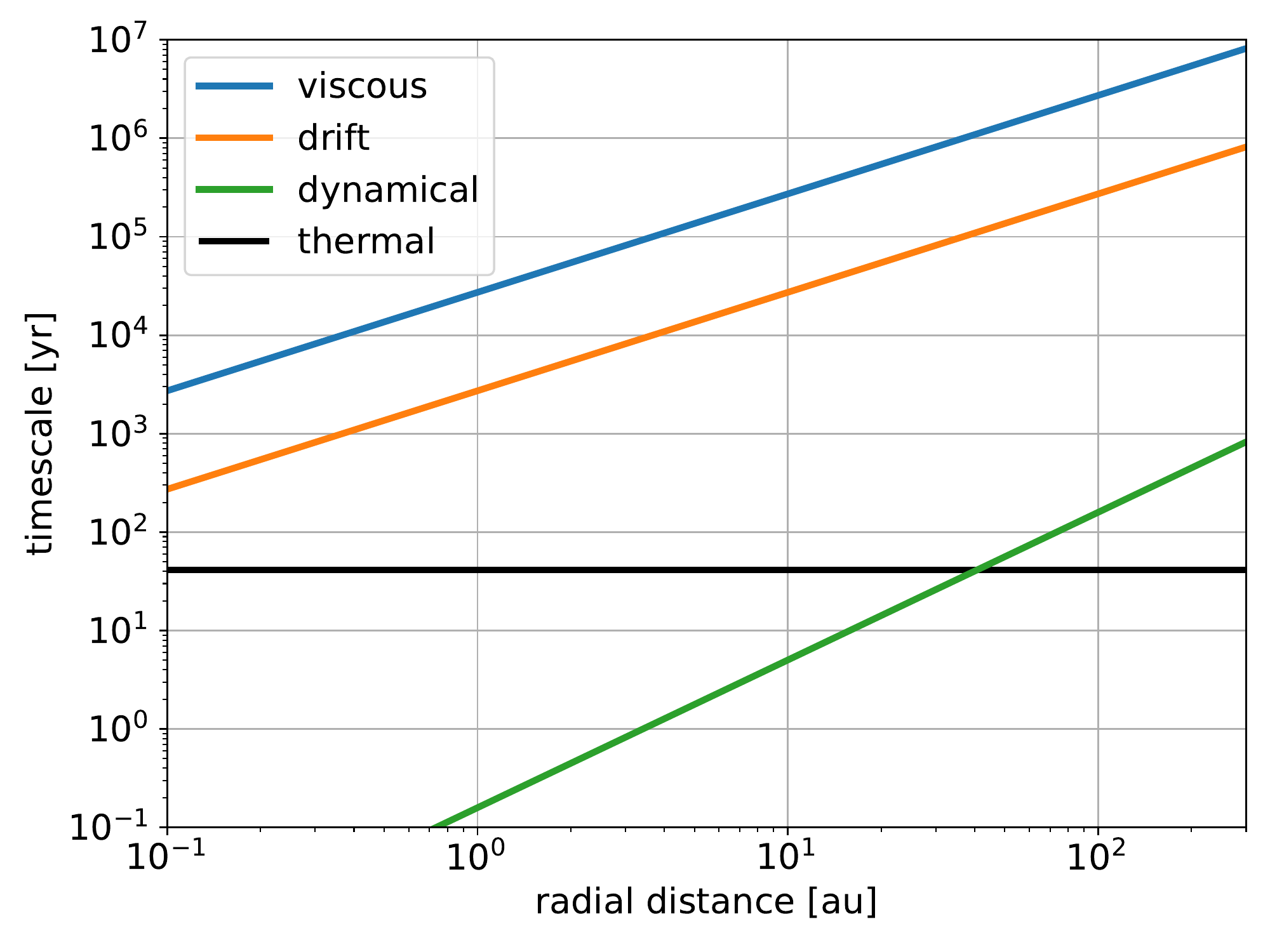}
\caption{
Comparison of the viscous evolution timescale (blue), dust radial drift timescale (orange), dynamical timescale (green) and the thermal timescale (black).
}
\label{fig:timescales}
\end{center}
\end{figure}

In this appendix, we compare the thermal timescale, viscous evolution timescale, dust radial drift timescale and disk dynamical timescale in Figure \ref{fig:timescales}.
The thermal timescale is 41.2 yr and independent on the radial distance for our setup  in the optically thick limit (Equation \ref{eq:t_th}).
The viscous evolution timescale $t_{\rm vis}$ is estimated as $t_{\rm vis}=r^{2}/\alpha c_{\rm s} h_{\rm g}$, where $\alpha$ is the viscosity parameter. 
We adopt $\alpha=10^{-2}$ and a simple power-law temperature profile of $150(r/{\rm au})^{-0.5}~{\rm K}$.
Using these, the viscous evolution timescale is estimated as $2.8\times10^{4}(r/{\rm au})~{\rm yr}$.
We clearly see that the viscous evolution timescale is much longer than the thermal timescale.
The radial drift timescale of dust grains $t_{\rm drift}$ is evaluated as $t_{\rm drift}=r/v_{\rm d}$, where $v_{\rm d}$ is the radial drift velocity.
The radial drift velocity is given as $\eta{\rm St}v_{\rm K}$, where $\eta=0.5(c_{\rm s}/v_{\rm K})^{2}{\rm dln~}p/{\rm dln~}r$ and ${\rm St}$ is the normalized stopping time of dust grains.
We adopt ${\rm St}=0.1$ and obtained  $t_{\rm drift}=2.8\times10^{3}(r/{\rm au})~{\rm yr}$
The dust radial drift timescale is much shorter than the viscous evolution timescale but much longer than the thermal timescale.
The dynamical timescale $t_{\rm dyn}$ is defined as $t_{\rm dyn}=\Omega_{\rm K}^{-1}=0.16(r/{\rm au})^{1.5}$.
The dynamical timescale is much shorter than the thermal timescale at $r\lesssim40~{\rm au}$ but longer at the outer region.
\label{sec:timescales}

\bibliographystyle{aasjournal}
\bibliography{reference}

\end{document}